A workflow for segmenting soil and plant X-ray CT images with deep learning in Google's Colaboratory

Devin A. Rippner[1,2], Pranav Raja[3], J. Mason Earles[3,4], Alexander Buchko[5], Mina Momayyezi[4], Fiona Duong[6], Dilworth Parkinson[7], Elizabeth Forrestel[4], Ken Shackel[8], Jeffrey Neyhart[9], and Andrew J. McElrone[4,10]

1. United States Department of Agriculture-Agricultural Research Service, Horticultural Crops Research Unit, Prosser, WA, USA

2. Department of Crop and Soil Sciences, Washington State University, Prosser, WA, USA

3. Department of Biological and Agricultural Engineering, University of California-Davis, Davis, CA, USA

4. Department of Viticulture and Enology, University of California-Davis, Davis, CA, USA

5. Department of Computer Science, California Polytechnic and State University, San Luis Obispo, CA, USA

6. Department of Integrative Biology, San Francisco State University, San Francisco, CA, USA

7. Advanced Light Source, Lawrence Berkeley National Laboratory, Berkeley, CA, USA

8. Department of Plant Sciences, University of California-Davis, Davis, CA, USA

9. United States Department of Agriculture-Agricultural Research Service, Genetic Improvement for Fruits & Vegetables Laboratory, Chatsworth, NJ, USA

10. United States Department of Agriculture-Agricultural Research Service, Crops Pathology and Genetics Research Unit, Davis, CA, USA

Abstract:
X-ray micro-computed tomography (X-ray μCT) has enabled the characterization of the properties and processes that take place in plants and soils at the micron scale. Despite the widespread use of this advanced technique, major limitations in both hardware and software limit the speed and accuracy of image processing and data analysis. Recent advances in machine learning, specifically the application of convolutional neural networks to image analysis, have enabled rapid and accurate segmentation of image data. Yet, challenges remain in applying convolutional neural networks to the analysis of environmentally and agriculturally relevant images. Specifically, there is a disconnect between the computer scientists and engineers, who build these AI/ML tools, and the potential end users in agricultural research, who may be unsure of how to apply these tools in their work. Additionally, the computing resources required for training and applying deep learning models are unique, more common to computer gaming systems or graphics design work, than to traditional computational systems. To navigate these challenges, we developed a modular workflow for applying convolutional neural networks to X-ray μCT images, using low-cost resources in Google's Colaboratory web application. Here we present the results of the workflow, illustrating how parameters can be optimized to achieve best results using example scans from walnut leaves, almond flower buds, and a soil aggregate. We expect that this framework will accelerate the adoption and use of emerging deep learning techniques within the plant and soil sciences.

**Introduction:**

Researchers have long been interested in analyzing the in-situ physical, chemical, and biological properties and processes that take place in plants and soils. To accomplish this, researchers have widely adopted the use of X-ray micro-computed tomography (X-ray µCT) for 3D analysis of flower buds, seeds, leaves, stems, roots, and soils (Anderson et al., 1990; Brodersen et al., 2010; Crestana et al., 1986, 1985; Cuneo et al., 2020; Duncan et al., 2022; Hapca et al., 2011; Helliwell et al., 2013; Mooney et al., 2012; Petrovic et al., 1982; Théroux-Rancourt et al., 2020; Xiao et al., 2021). In plants, researchers have used X-ray µCT to visualize the internal structures of leaves, allowing for the quantification of $CO_2$ diffusion through the leaf based on path length tortuosity from the stomata to the mesophyll (Mathers et al., 2018; Théroux-Rancourt et al., 2021). Other applications of X-ray µCT in plants include the visualization of embolism formation and repair in plant xylem tissue, allowing for the development of new models to better understand drought stress recovery, along with non-destructive quantification of carbohydrates in plant stems (Brodersen et al., 2010; Earles et al., 2018; Torres-Ruiz et al., 2015). In soils, X-ray µCT was used to visualize soil porosity, soil aggregate distribution, and plant root growth (Ahmed et al., 2016; Gerth et al., 2021; Helliwell et al., 2013; Keyes et al., 2022.; Mairhofer et al., 2013; Mooney et al., 2012; Tracy et al., 2010; Yudina and Kuzyakov, 2019). Despite the wide use of advanced imaging techniques like X-ray µCT and imaging more generally in agricultural research, major limitations in both hardware and software hinder the speed and accuracy of image processing and data analysis.

Historically X-ray µCT data collection was extremely time consuming, and resource intensive as individual scans can exceed 50 Gb in size. Data acquisition rates were limited by the ability of X-ray detectors to transfer data to computers, limited hard drive storage capacity once the data was transferred, and intensive hardware requirements that limited the size of files that could be analyzed at any given time. Many of these constraints have been removed as detector hardware has improved, hard drive storage transfer speed and space has increased, and computing hardware has advanced. Now a major limiting step to the widespread use of X-ray μCT in the agricultural sciences is data analysis; while data can be acquired in hours to seconds, the laborious task of hand segmenting images can lead to analysis times of weeks to years for large data sets (Théroux-Rancourt et al., 2020).

Recent advances in machine learning, specifically the application of convolutional neural networks to image analysis, have enabled rapid and accurate segmentation of image data (Chen et al., 2018; Long et al., 2015; Raja et al., 2021; Ronneberger et al., 2015; Smith et al., 2020; von Chamier et al., 2021). Such applications have met with great success in medical imaging analysis, outperforming radiologists for early cancer diagnosis in X-ray μCT images (Lotter et al., 2021). However, challenges remain to applying convolutional neural networks to the analysis of agriculturally relevant X-ray μCT images. Specifically, training accurate models for image segmentation requires the production of hand annotated training datasets, which is time consuming and requires specialized expertise to properly annotate training image data (Kamilaris and Prenafeta-Boldú, 2018). Further, the computing resources required for training and applying deep learning models are unique, more common to computer gaming systems or graphics design work, rather than traditional computational systems (Gao et al., 2020; Ofori et al., 2022).

To navigate these challenges, we developed a modular workflow for image annotation and segmentation using open-source tools to empower scientists that use X-ray μCT in their work. Specifically, image annotation is done in ImageJ; while this does not prevent the need for experts to annotate images, it does allow experts to annotate their images without using proprietary software. The semantic segmentation of X-ray µCT image data is accomplished using Google's Colab to run PyTorch implementations of a Fully Convolutional Network (FCN) with a ResNet-101 backbone (Chen et al., 2017; He et al., 2015; Long et al., 2015; Paszke et al., 2019; Ronneberger et al., 2015). The FCN architecture, while older, allows for model development on variable size images due to the exclusion of fully connected layers in the FCN architecture (Long et al., 2015; Ronneberger et al., 2015). In addition to X-ray µCT datasets, the workflow is flexible enough to work on virtually any image dataset, as long as corresponding annotated images are available for model training. Additionally, by developing and deploying the code in Google's Colaboratory, users have access to free or low-cost GPU resources that might otherwise be cost prohibitive to access (Rippner et al., 2022). If users have access to better hardware than is available through Colaboratory, the notebooks can be run locally to utilize advanced hardware. Additional code is also available to use this workflow on high performance computing systems using batch scheduling (Rippner et al., 2022). This method for analyzing X-ray µCT data allows users to rapidly extract information on important biological, chemical, and physical processes that occur in plants and soils from complex datasets without the need to learn to code extensively or invest in expensive computational hardware.

**Materials and Methods:**

In the following section we will describe the parameters under which our CT data was collected, how the CT image data was annotated for model training, and the parameters used to train the various models. The actual workflow and corresponding training video can be found on Github (Rippner et al., 2022). For reproducibility purposes, the data sets used for training the models featured in this paper can be found a repository hosted by the United States Department of Agriculture, National Agricultural Library (Rippner et al., 2022).

CT data Acquisition:

Six individual leaf sections (3 mm x 7 mm) from 6 unique accessions of English walnuts (*Juglans regia)* and an air dried soil aggregate collected from the top 15 cm of a Yolo silt loam (Fine-silty, mixed, superactive, nonacid, thermic Mollic Xerofluvents) at the UC Davis Russel Ranch Sustainable Agricultural Facility were scanned at 23 keV using the 10× objective lens with a pixel resolution of 650 nanometers on the X-ray µCT beamline (8.3.2) at the Advanced Light Source (ALS) in Lawrence Berkeley National Laboratory (LBNL), Berkeley, CA USA. Additionally, an almond flower bud (*Prunis dulcis*) was scanned using a 4× lens with a pixel resolution of 1.72 µm on the same beamline. Raw tomographic image data was reconstructed using the TomoPy tomographic image reconstruction engine (Gürsoy et al., 2014). Reconstructions were converted to 8-bit tif or png format using ImageJ or the PIL package in Python before further processing (Fig. 1) (Kemenade et al., 2022; Schindelin et al., 2012).

Image Annotation:

Leaf images were annotated in ImageJ following Théroux-Rancourt et al. (2020) (Fig.1) Flower bud and soil aggregate images were annotated using Intel's Computer Vision Annotation Tool (CVAT) and ImageJ (Fig. 1) (Schindelin et al., 2012). Both CVAT and ImageJ are free to use and open source. To annotate the flower bud and soil aggregate, images were imported into CVAT. The exterior border of the bud (i.e. bud scales) and flower were annotated in CVAT and exported as masks. Similarly, the exterior of the soil aggregate and particulate organic matter identified by eye were annotated in CVAT and exported as masks. To annotate air spaces in both the bud and soil aggregate, images were imported into ImageJ. A gaussian blur was applied to the image to decrease noise and then the air space was segmented using thresholding. After applying the threshold, the selected air space region was converted to a binary image with white representing the air space and black representing everything else. This binary image was overlaid upon the original image and the air space within the flower bud and aggregate was selected using the "free hand" tool. Air space outside of the region of interest for both image sets was eliminated. The quality of the air space annotation was then visually inspected for accuracy against the underlying original image; incomplete annotations were corrected using the brush or pencil tool to paint missing air space white and incorrectly identified air space black. Once the annotation was satisfactorily corrected, the binary image of the air space was saved. Finally, the annotations of the bud and flower or aggregate and organic matter were opened in ImageJ and the associated air space mask was overlaid on top of them forming a three-layer mask suitable for training the fully convolutional network.

Training General *Juglans* Leaf Segmentation Model:

Images and associated annotations from 6 walnut leaf scans were uploaded to Google Drive (Fig 1). Using Google's Colaboratory resources, a PyTorch implementation of a Fully Convolutional Network (FCN) with a ResNet-101 backbone was used to train 10 models using 5 image/annotation pairs from 1, 2, 3, 4, and 5 leaves; (5, 10, 15, 20, and 25 images/annotation pairs respectively) (Fig. 1; SI Table 1, 2; SI Fig. 1, 2) (He et al., 2015; Paszke et al., 2019). Models pre-trained on the COCO train2017 dataset were imported and the original classifier was substituted for a new classifier based on 6 potential pixel classes: background, epidermis, mesophyll tissue, air space, bundle sheath extension tissue, or vein tissue. The pre-trained model weights were modified using an Adam optimizer for stochastic optimization with the learning rate set to 0.001 and a binary cross-entropy loss function (Kingma and Ba, 2017). To help avoid overfitting the training data, the data was augmented using Albumentations package in Python to flip and rotate a subset of the images during model training (Buslaev et al., 2020). Half of the image/annotation pairs were used for training and half were used for validation of the model during training. The batch size was set at 1 for training due to graphics processing unit (GPU) constraints in Colaboratory. A mixture of NVIDA T4, P100, V100 GPUs were used for training depending on the allocation assigned by Google Cloud Services. Such GPU's are available when using the free version of Google's Colaboratory, or the low cost ($9.99/month) subscription based Colaboratory Pro (Mountain View, CA, USA). A benefit of limiting the batch size to 1 was the ability to train on variably sized images.

The accuracy, precision, recall, and F1 score of these models were calculated after testing on 5 images from the 6th leaf that was not involved in training or validation of the generated models in any way. In our work accuracy, precision, recall, and f1 scores are defined as:

*Equation 1:*

$$\text{Accuracy} = \frac{TP+TN}{TP+TN+FP+FN}$$

*Equation 2:*

$$\text{Precision} = \frac{(TP+1E-9)}{(TP+FP+1E-9)}$$

*Equation 3:*

$$\text{Recall} = \frac{(TP+1E-9)}{(TP+FN+1E-9)}$$

*Equation 4:*

$$F1 = \frac{(TP+1E-9)}{(TP+1E-9)+\frac{1}{2}(FP+FN)}$$

Where TP = true positive prediction on a pixelwise basis, FP = false positive prediction on a pixelwise basis, TN = true negative prediction on a pixelwise basis, and FN = false negative prediction on a pixelwise basis. A correction factor of 1E-9 was included in equations 2-4 to prevent *Not a Number* errors in python when the denominator of the equations is 0 due to the lack of TP, FP, or FN values when no prediction is made for a non-existent material class in a particular image (Powers, 2020).

The evaluation results for each of the 10 models generated after training on 1, 2, 3, 4, and 5 leaves were compiled using the Panda's library in Python and visualized using the Seaborn library (SI Table 1) (McKinney, 2011; Oliphant, 2007; Waskom, 2021).

The number of training epochs (i.e. iterative learning passes through the complete data set) to train a satisfactory model was also evaluated using the 25 image/annotation pairs taken from 5 annotated leaves. This number of training images was found to give the best results with a fixed number of epochs for model training. Ten models were generated after training for 10, 25, 50, 100, and 200 epochs. The accuracy, precision, recall, and F1 scores for these models were calculated after evaluating the same 5 images from the 6$^{th}$ leaf that was not used for model training and validation. Binary image outputs for each material type generated for the leaves were stacked and rendered in 3-dimensions using ORS Dragonfly (Object Research Systems, Montréal, Canada).

Training Models for Segmenting Flower Buds and Soil Aggregates:

A mixture of NVIDA T4, P100, V100, and A100 GPUs were used for training models for segmenting an almond flower bud and a soil aggregate. Such GPU's are available when using the free version of Google's Colaboratory, the low cost ($9.99/month) subscription based

Colaboratory Pro, or the higher cost ($49.99) subscription based Colaboratory Pro + (Mountain View, CA, USA). Training and validation images for both the almond flower bud and the soil aggregate had to be downscaled to 50% size in the x and y dimensions to fit on the T4, P100, and V100 video cards due to VRAM limitations (16 Gb VRAM) (SI Table 1). When using the A100 GPU (40 Gb VRAM) available through Colaboratory Pro +, images used for model training and validation from the flower bud and soil aggregate were only downscaled to 85% in the x and y dimensions, representing a large gain in image data for training and validation (SI Table 1). Again, models pre-trained on the COCO train2017 dataset were imported and the original classifier was substituted for a new classifier based on 4 potential pixel classes. For the soil aggregate, these were background, mineral solids, pore space and particulate organic matter; for the almond bud there were background, bud scales, leaf tissues, and air space. The pre-trained model weights were modified using an Adam optimizer for stochastic optimization with the learning rate set to 0.001 and a binary cross-entropy loss function (Kingma and Ba, 2017). To help avoid overfitting the training data, the data was augmented using Albumentations package in Python to flip and rotate a subset of the images during model training (Buslaev et al., 2020). Models were trained for 200 epochs as model loss for these data was previously found to plateau between 100 and 200 epochs. Model accuracy, precision, recall and F1 scores were calculated as above after testing on 5 independently annotated images from the same flower bud and soil aggregate that were not used for model training or validation. Binary image outputs for each material type generated for the almond bud and soil aggregate were stacked and rendered in 3d using ORS Dragonfly (Object Research Systems, Montéal, Canada).

Results:

Accuracy, precision, recall, and F1 scores for a general model to identify and segment specific walnut leaf tissues on a pixel-wise basis plateaued after training and validation on at least 3 annotated leaves. For epidermis and mesophyll tissues, prediction F1 scores were >80% while prediction F1 scores were generally > 75% for bundle sheath extensions and >70% for airspaces. The lowest prediction F1 scores were achieved for veins tissues (~60%) and the highest for the background class (>95%) (Fig. 2, Fig. 3). Precision scores were generally higher than recall scores except in the case of vein tissue identification, where the models tended to over-predict the occurrence of the vein tissue class. F1 score variability across all prediction classes decreased as the number of leaves used during training and validation increased. Model F1 score variability was highest for the bundle sheath extension and vein tissue classes, likely due to colocation of the two tissue classes. When evaluating generalized walnut leaf model performance with increasing training epochs, model F1 scores plateaued after 50 epochs. Model F1 score variability was consistent after 50 epochs, with no improvement after additional training time (Fig 4, Fig 5). This was particularly true for the vein tissue class which took the most training for consistent identification.

Model performance for the segmentation of an almond flower bud and a soil aggregate was hindered by the downscaling necessary to fit the training and validation data on video cards with 16 Gb of VRAM. When downscaled to 0.5 size in the x and y dimensions (25% size), the best model F1 scores were 99.8%, 99.1%, 92.4%, and 71.5% for the background, bud scale, flower, and air space classes, respectively. At the same scaling for the soil aggregate, the best model F1 scores were 99.4%, 83.7%, 55.6%, and 71.3% for the background, solid, pore, and organic matter classes respectively. With access to the A100 GPU with 40 Gb of VRAM, training images were only downscaled to 0.85 size in the x and y dimensions (72% size). This yielded best model F1 scores of 99.9%, 99.4%, 94.5%, and 74.4% for the background, bud scale, flower, and air space classes, respectively. At the same scaling for the soil aggregate the best model F1 scores were 99.4%, 91.3%, 76.7%, and 74.9% for the background, solid, pore, and organic matter classes respectively. While the improvements in model performance with increased scaling were modest for the flower bud, they were large for the soil aggregate, likely due to the intricacy of the aggregate that was lost with downscaling (Fig. 6).

Model outputs are binary 2-dimensional data from which information like material area, perimeter or other morphological traits can be extracted using downstream image analysis functions. These data are saved as sequences of arrays which allows for the 3d visualization of the segmented materials (Fig 7). Additionally, 3d data can be extracted from the array sequences using Python libraries like Numpy or using other programming languages like R or Matlab.

**Discussion:**

Generalized models for image segmentation are typically generated after training on hundreds to millions of images (Belthangady and Royer, 2019; Khened et al., 2021; O'Mahony et al., 2020; Shahinfar et al., 2020). Due to the limited availability of leaf, bud, or soil x-ray CT scans, such a training image library simply doesn't exist (Moen et al., 2019). However, we found that we can generate accurate models using 5 annotated slices from at least 3 unique leaf scans. This discrepancy is likely the result of the consistent image collection settings (resolution) on the same imaging platform (Beamline 8.3.2) which simplifies the learning process (Fei et al., 2021; Silwal et al., 2021). Our results are comparable to those previously achieved by Théroux-Rancourt et al. (2020) on X-ray CT images of plant leaves. That method, based on random forest classification, requires hand annotating 6 images from every single scan and is designed for extracting data from leaf X-ray CT images exclusively. Specifically, the precision and recall scores for the background (>95%), mesophyll tissue (>80%), epidermis tissue (>80%), and bundle sheath extension (>75%) classes were equal to those achieved by Théroux-Rancourt et al. (2020). The current approach had lower precision and recall scores (>75%) for air space identification compared to Théroux-Rancourt et al. (2020) (>90%), but higher precision and recall scores for vein tissue identification (>60% vs <55%, respectively). Despite similarities in the quality of results, the current method decreases segmentation time from hours to minutes compared to Théroux-Rancourt et al. (2020) and can be applied to any X-ray CT image data set.

Our training batch size was limited to 1 by a combination of factors including variably sized training images for the leaf scans and hardware limitations for the almond bud and soil aggregate scans. Typically, batch size selection is constrained by a combination of hardware and the number of images used for training; the smaller the batch size, the longer training takes (Smith et al., 2017). This presents a significant barrier when training using millions of images but is not an issue when only tens of training images are available.

Epoch selection is an important component of maximizing model accuracy, precision, recall, and F1 scores. Training models for too few epochs leads to substandard model performance while over training with too many epochs wastes time and can lead to overfitting (Baldeon Calisto and Lai-Yuen, 2020; Li et al., 2019; Pan et al., 2020). Typically models trained with small batch sizes require more training epochs for satisfactory performance compared to models trained on large batch sizes (Smith, 2018). This can greatly increase training times if many images are required for training to achieve satisfactory model performance (von Chamier et al., 2021).

Beyond batch size and epoch number, image size plays a significant role in model performance. Large images take up more GPU memory than smaller images (Sabottke and Spieler, 2020). GPU memory can be conserved by decreasing batch size, but once batch size has been reduced to 1, the only option is to downscale images for training. However, this results in a significant loss of information in the images, hindering model performance as fine details are lost (Sabottke and Spieler, 2020). This was particularly problematic with the soil aggregate scans which contained fine pore spaces which were lost when the images were downscaled to 0.5 in the x and y dimensions to fit on GPUs with 16 Gb of VRAM. Downscaling to this degree results in a loss of 75% of the image information. Only when GPUs with more VRAM were used could images be downscaled less, resulting in improved performance for models trained on these higher resolution images.

By stacking the sequences of data arrays produced by the model, novel information can be gained about processes that occur over 3 dimensions. Taking this approach, Théroux-Rancourt et al. (2017) previously showed that mesophyll surface area exposed to intercellular air space is underestimated when using 2D rather than 3D approaches. Similarly, Trueba et al. (2021) found that the 3D organization of leaf tissues had a direct impact on plant water use and carbon uptake. In soils, it well understood that pore tortuosity plays a key role in understanding processes like water infiltration and $O_2/CO_2$ diffusion. As Baveye et al. (2018) highlighted in their review, the rapid segmentation of soil X-ray μCT data has long been a major hurdle to understanding these processes. Our workflow simplifies and accelerates this process, enabling researchers to rapidly extract information from their X-ray μCT data. Our approach is similar to those developed by von Chamier et al. (2021) and Smith et al. (2020), but is more flexible as it works with variably sized images and allows for multi-label semantic segmentation.

**Conclusions:**

With the current work, we present a workflow for using open-source software generate models to segment X-ray μCT images. These models can be specific to an image set (segmenting a single soil aggregate or almond bud) or be generalized for a specific use case such as segmenting leaf scans. We demonstrated that a limited number of annotated images can achieve satisfactory results without excessively long training time. The workflow can be run locally, in Google's Colaboratory Notebook, or adapted for use on high performance computing platforms. By using GPU resources, the rate of segmentation can be dramatically increased, taking less than 0.02 seconds per image. This allows users to segment scans in minutes, a significant speed gain compared to other methods with similar precision and recall (often >90%) across a variety of sample scans (Arganda-Carreras et al., 2017; Théroux-Rancourt et al., 2020). This will allow researchers to gain novel insights into the role that 3d architecture of soil and plant samples plays in a variety of important processes.

Figure 1: A schematic of the segmentation workflow from image reconstruction, image annotation, model training, model use, and data extraction. Blue indicates a process that is done at the instrumentation site, green is a process done on local computers using a subset of the data in ImageJ or CVAT, purple indicates a process done in Google's Colaboratory, on a high-performance computing cluster, or locally, and yellow indicates a process that can be done with the users software of choice including Python, R, or Matlab.

Figure 2: Accuracy, precision, recall, and F1 scores as a function of uniquely annotated leaf number (5 annotated images per leaf) used to predict tissue classes in X-ray µCT images from an independent leaf on which the models were not trained or validated. Circles represent unique predictions from 10 uniquely generated models per leaf number; dark gray lines represent the mean value of the 10 models while thick light gray lines represent the 95% confidence interval of the values.

Figure 3: Visual representation of the model outputs for a single walnut leaf image; top image is a X-ray CT scan taken from a leaf that was not used for training or validation of the applied models; next is the hand annotated image of the scan followed by the outputs from the best performing model trained on 1, 3, and 5 leaves, respectively. For the walnut leaf segmentations the background is light gray, the epidermis is dark gray, the mesophyll is black, the air space is white, the bundle sheath extensions are middle gray, and the veins are lightest gray.

Figure 4: Accuracy, precision, recall, and F1 scores as a function increasing epoch number for models trained on annotated images from 5 leaves (5 annotated images per leaf) used to predict tissue classes from X-ray µCT images from an independent leaf on which the models were not trained or validated. Circles represent unique predictions from 10 uniquely generated models per

epoch number; dark gray lines represent the mean value of the 10 models for each tissue type while thick light gray lines represent the 95% confidence interval of the values.

Figure 5: Visual representation of the model outputs for a single walnut leaf image cross-section; top image is a X-ray CT scan taken from a leaf that was not used for training or validation of the applied models; next is the hand annotated image of the scan followed by the outputs from the best performing model trained for 10, 100, and 200 epochs respectively. For the walnut leaf segmentation, the background is light gray, the epidermis is dark gray, the mesophyll is black, the air space is white, the bundle sheath extensions are middle gray, and the veins are lightest gray.

Figure 6: Visual representation of the model outputs for an almond flower bud and a soil aggregate; top images are X-ray CT scans of the flower bud and soil aggregate; next are the hand annotated images of the scans followed by the outputs from the best performing models trained at 0.5 and 0.85 scale, respectively. For the almond flower bud, background is black, bud scales are dark gray, flower tissue is light gray, and air spaces are white. For the soil aggregate, background is black, mineral solids are dark gray, pore spaces are white, and organic matter is light gray.

Figure 7: 3D visual representation of the stacked model outputs for a walnut leaf, a almond flower bud, and a soil aggregate; top images are X-ray CT scans of the bud and soil aggregate; next are the hand. In the walnut leaf images, the background is black, the epidermis is dark green, the mesophyll is light green, the air space is white, the bundle sheath extensions burnt orange, and the veins are brown. In the almond bud images, background is black, bud scales are brown, flower tissue is pink, and air spaces are white. For the soil aggregate, background is black, mineral solids are dark gray, pore spaces are blue, and organic matter is brown.

References:

Ahmed, S., Klassen, T.N., Keyes, S., Daly, M., Jones, D.L., Mavrogordato, M., Sinclair, I., Roose, T., 2016. Imaging the interaction of roots and phosphate fertiliser granules using 4D X-ray tomography. Plant and Soil 401, 125–134.

Anderson, S.H., Peyton, R.L., Gantzer, C.J., 1990. Evaluation of constructed and natural soil macropores using X-ray computed tomography. Geoderma 46, 13–29.

Arganda-Carreras, I., Kaynig, V., Rueden, C., Eliceiri, K.W., Schindelin, J., Cardona, A., Sebastian Seung, H., 2017. Trainable Weka Segmentation: a machine learning tool for microscopy pixel classification. Bioinformatics 33, 2424–2426. https://doi.org/10.1093/bioinformatics/btx180

Baldeon Calisto, M., Lai-Yuen, S.K., 2020. AdaEn-Net: An ensemble of adaptive 2D–3D Fully Convolutional Networks for medical image segmentation. Neural Networks 126, 76–94. https://doi.org/10.1016/j.neunet.2020.03.007

Belthangady, C., Royer, L.A., 2019. Applications, promises, and pitfalls of deep learning for fluorescence image reconstruction. Nat Methods 16, 1215–1225. https://doi.org/10.1038/s41592-019-0458-z

Brodersen, C.R., McElrone, A.J., Choat, B., Matthews, M.A., Shackel, K.A., 2010. The dynamics of embolism repair in xylem: in vivo visualizations using high-resolution computed tomography. Plant physiology 154, 1088–1095.

Buslaev, A., Iglovikov, V.I., Khvedchenya, E., Parinov, A., Druzhinin, M., Kalinin, A.A., 2020. Albumentations: fast and flexible image augmentations. Information 11, 125.

Chen, L.-C., Papandreou, G., Schroff, F., Adam, H., 2017. Rethinking atrous convolution for semantic image segmentation. arXiv preprint arXiv:1706.05587.

Chen, L.-C., Zhu, Y., Papandreou, G., Schroff, F., Adam, H., 2018. Encoder-decoder with atrous separable convolution for semantic image segmentation, in: Proceedings of the European Conference on Computer Vision (ECCV). pp. 801–818.

Crestana, S., Cesareo, R., Mascarenhas, S., 1986. Using a computed tomography miniscanner in soil science. Soil Science 142, 56.

Crestana, S., Mascarenhas, S., Pozzi-Mucelli, R.S., 1985. Static and dynamic three-dimensional studies of water in soil using computed tomographic scanning1. Soil Science 140, 326–332.

Cuneo, I.F., Barrios-Masias, F., Knipfer, T., Uretsky, J., Reyes, C., Lenain, P., Brodersen, C.R., Walker, M.A., McElrone, A.J., 2020. Differences in grapevine rootstock sensitivity and recovery from drought are linked to fine root cortical lacunae and root tip function. New Phytologist.

Duncan, K.E., Czymmek, K.J., Jiang, N., Thies, A.C., Topp, C.N., 2022. X-ray microscopy enables multiscale high-resolution 3D imaging of plant cells, tissues, and organs. Plant Physiology 188, 831–845. https://doi.org/10.1093/plphys/kiab405

Earles, J.M., Knipfer, T., Tixier, A., Orozco, J., Reyes, C., Zwieniecki, M.A., Brodersen, C.R., McElrone, A.J., 2018. In vivo quantification of plant starch reserves at micrometer resolution using X-ray microCT imaging and machine learning. New Phytologist 218, 1260–1269. https://doi.org/10.1111/nph.15068

Fei, Z., Olenskyj, A.G., Bailey, B.N., Earles, M., 2021. Enlisting 3D Crop Models and GANs for More Data Efficient and Generalizable Fruit Detection. Presented at the Proceedings of the IEEE/CVF International Conference on Computer Vision, pp. 1269–1277.

Gao, Y., Liu, Y., Zhang, H., Li, Z., Zhu, Y., Lin, H., Yang, M., 2020. Estimating GPU memory consumption of deep learning models, in: Proceedings of the 28th ACM Joint Meeting on European Software Engineering Conference and Symposium on the Foundations of Software Engineering. Association for Computing Machinery, New York, NY, USA, pp. 1342–1352.

Gerth, S., Claußen, J., Eggert, A., Wörlein, N., Waininger, M., Wittenberg, T., Uhlmann, N., 2021. Semiautomated 3D root segmentation and evaluation based on X-ray CT imagery. Plant Phenomics 2021.

Gürsoy, D., De Carlo, F., Xiao, X., Jacobsen, C., 2014. TomoPy: a framework for the analysis of synchrotron tomographic data. Journal of synchrotron radiation 21, 1188–1193.

Hapca, S.M., Wang, Z.X., Otten, W., Wilson, C., Baveye, P.C., 2011. Automated statistical method to align 2D chemical maps with 3D X-ray computed micro-tomographic images of soils. Geoderma 164, 146–154. https://doi.org/10.1016/j.geoderma.2011.05.018

He, K., Zhang, X., Ren, S., Sun, J., 2015. Deep Residual Learning for Image Recognition. arXiv:1512.03385 [cs].

Helliwell, J.R., Sturrock, C.J., Grayling, K.M., Tracy, S.R., Flavel, R.J., Young, I.M., Whalley, W.R., Mooney, S.J., 2013. Applications of X-ray computed tomography for examining biophysical interactions and structural development in soil systems: a review. European Journal of Soil Science 64, 279–297.

Kamilaris, A., Prenafeta-Boldú, F.X., 2018. Deep learning in agriculture: A survey. Computers and Electronics in Agriculture 147, 70–90. https://doi.org/10.1016/j.compag.2018.02.016

Kemenade, H. van, Murray, A., wiredfool, Clark, A., Karpinsky, A., Baranovič, O., Gohlke, C., Dufresne, J., Schmidt, D., Kopachev, K., Houghton, A., Mani, S., Landey, S., vashek,

Ware, J., Douglas, J., T, S., Caro, D., Martinez, U., Kossouho, S., Lahd, R., Lee, A., Brown, E.W., Tonnhofer, O., Bonfill, M., Rowlands (변기호), P., Al-Saidi, F., Novikov, G., Górny, M., 2022. python-pillow/Pillow: 9.0.1. Zenodo. https://doi.org/10.5281/zenodo.5953590

Keyes, S., van Veelen, A., McKay Fletcher, D., Scotson, C., Koebernick, N., Petroselli, C., Williams, K., Ruiz, S., Cooper, L., Mayon, R., Duncan, S., Dumont, M., Jakobsen, I., Oldroyd, G., Tkacz, A., Poole, P., Mosselmans, F., Borca, C., Huthwelker, T., Jones, D.L., Roose, T., n.d. Multimodal correlative imaging and modelling of phosphorus uptake from soil by hyphae of mycorrhizal fungi. New Phytologist n/a. https://doi.org/10.1111/nph.17980

Khened, M., Kori, A., Rajkumar, H., Krishnamurthi, G., Srinivasan, B., 2021. A generalized deep learning framework for whole-slide image segmentation and analysis. Sci Rep 11, 11579. https://doi.org/10.1038/s41598-021-90444-8

Kingma, D.P., Ba, J., 2017. Adam: A Method for Stochastic Optimization. arXiv:1412.6980 [cs].

Li, H., Li, J., Guan, X., Liang, B., Lai, Y., Luo, X., 2019. Research on Overfitting of Deep Learning, in: 2019 15th International Conference on Computational Intelligence and Security (CIS). Presented at the 2019 15th International Conference on Computational Intelligence and Security (CIS), pp. 78–81. https://doi.org/10.1109/CIS.2019.00025

Long, J., Shelhamer, E., Darrell, T., 2015. Fully convolutional networks for semantic segmentation, in: Proceedings of the IEEE Conference on Computer Vision and Pattern Recognition. pp. 3431–3440.

Lotter, W., Diab, A.R., Haslam, B., Kim, J.G., Grisot, G., Wu, E., Wu, K., Onieva, J.O., Boyer, Y., Boxerman, J.L., Wang, M., Bandler, M., Vijayaraghavan, G.R., Gregory Sorensen,

A., 2021. Robust breast cancer detection in mammography and digital breast tomosynthesis using an annotation-efficient deep learning approach. Nature Medicine 27, 244–249. https://doi.org/10.1038/s41591-020-01174-9

Mairhofer, S., Zappala, S., Tracy, S., Sturrock, C., Bennett, M.J., Mooney, S.J., Pridmore, T.P., 2013. Recovering complete plant root system architectures from soil via X-ray μ-computed tomography. Plant methods 9, 1–7.

Mathers, A.W., Hepworth, C., Baillie, A.L., Sloan, J., Jones, H., Lundgren, M., Fleming, A.J., Mooney, S.J., Sturrock, C.J., 2018. Investigating the microstructure of plant leaves in 3D with lab-based X-ray computed tomography. Plant methods 14, 1–12.

McKinney, W., 2011. pandas: a foundational Python library for data analysis and statistics. Python for High Performance and Scientific Computing 14, 1–9.

Moen, E., Bannon, D., Kudo, T., Graf, W., Covert, M., Van Valen, D., 2019. Deep learning for cellular image analysis. Nat Methods 16, 1233–1246. https://doi.org/10.1038/s41592-019-0403-1

Mooney, S.J., Pridmore, T.P., Helliwell, J., Bennett, M.J., 2012. Developing X-ray computed tomography to non-invasively image 3-D root systems architecture in soil. Plant and soil 352, 1–22.

Ofori, M., El-Gayar, O., O'Brien, A., Noteboom, C., 2022. A Deep Learning Model Compression and Ensemble Approach for Weed Detection, in: Proceedings of the 55th Hawaii International Conference on System Sciences.

Oliphant, T.E., 2007. Python for scientific computing. Computing in Science & Engineering 9, 10–20.

O'Mahony, N., Campbell, S., Carvalho, A., Harapanahalli, S., Hernandez, G.V., Krpalkova, L., Riordan, D., Walsh, J., 2020. Deep Learning vs. Traditional Computer Vision, in: Arai, K., Kapoor, S. (Eds.), Advances in Computer Vision, Advances in Intelligent Systems and Computing. Springer International Publishing, Cham, pp. 128–144. https://doi.org/10.1007/978-3-030-17795-9_10

Pan, Z., Emaru, T., Ravankar, A., Kobayashi, Y., 2020. Applying Semantic Segmentation to Autonomous Cars in the Snowy Environment. arXiv:2007.12869 [cs].

Paszke, A., Gross, S., Massa, F., Lerer, A., Bradbury, J., Chanan, G., Killeen, T., Lin, Z., Gimelshein, N., Antiga, L., 2019. Pytorch: An imperative style, high-performance deep learning library. arXiv preprint arXiv:1912.01703.

Petrovic, A.M., Siebert, J.E., Rieke, P.E., 1982. Soil bulk density analysis in three dimensions by computed tomographic scanning. Soil Science Society of America Journal 46, 445–450.

Powers, D.M., 2020. Evaluation: from precision, recall and F-measure to ROC, informedness, markedness and correlation. arXiv preprint arXiv:2010.16061.

Raja, P., Olenskyj, A., Kamangir, H., Earles, M., 2021. Simultaneously Predicting Multiple Plant Traits from Multiple Sensors via Deformable CNN Regression.

Rippner, D., Momayyezi, M., Shackel, K., Raja, P., Buchko, A., Duong, F., Parkinson, D., Earles, J., Forrestel, E., McElrone, A., 2022. X-ray CT data with semantic annotations for the paper "A workflow for segmenting soil and plant X-ray CT images with deep learning in Google's Colaboratory." https://doi.org/10.15482/USDA.ADC/1524793

Rippner, D., Pranav, R., Earles, J.M., Buchko, A., Neyhart, J., 2022. Welcome to the workflow associated with the manuscript: A workflow for segmenting soil and plant X-ray CT images with deep learning in Google's Colaboratory.

Ronneberger, O., Fischer, P., Brox, T., 2015. U-net: Convolutional networks for biomedical image segmentation, in: International Conference on Medical Image Computing and Computer-Assisted Intervention. Springer, pp. 234–241.

Sabottke, C.F., Spieler, B.M., 2020. The Effect of Image Resolution on Deep Learning in Radiography. Radiology: Artificial Intelligence 2, e190015. https://doi.org/10.1148/ryai.2019190015

Schindelin, J., Arganda-Carreras, I., Frise, E., Kaynig, V., Longair, M., Pietzsch, T., Preibisch, S., Rueden, C., Saalfeld, S., Schmid, B., Tinevez, J.-Y., White, D.J., Hartenstein, V., Eliceiri, K., Tomancak, P., Cardona, A., 2012. Fiji: an open-source platform for biological-image analysis. Nat Methods 9, 676–682. https://doi.org/10.1038/nmeth.2019

Shahinfar, S., Meek, P., Falzon, G., 2020. "How many images do I need?" Understanding how sample size per class affects deep learning model performance metrics for balanced designs in autonomous wildlife monitoring. Ecological Informatics 57, 101085. https://doi.org/10.1016/j.ecoinf.2020.101085

Silwal, A., Parhar, T., Yandun, F., Kantor, G., 2021. A Robust Illumination-Invariant Camera System for Agricultural Applications. arXiv:2101.02190 [cs].

Smith, A.G., Han, E., Petersen, J., Olsen, N.A.F., Giese, C., Athmann, M., Dresbøll, D.B., Thorup-Kristensen, K., 2020. RootPainter: deep learning segmentation of biological images with corrective annotation. Bioarxiv NA-NA.

Smith, L.N., 2018. A disciplined approach to neural network hyper-parameters: Part 1 -- learning rate, batch size, momentum, and weight decay. arXiv:1803.09820 [cs, stat].

Smith, S.L., Kindermans, P.-J., Ying, C., Le, Q.V., 2017. Don't decay the learning rate, increase the batch size. arXiv preprint arXiv:1711.00489.

Théroux-Rancourt, G., Jenkins, M.R., Brodersen, C.R., McElrone, A., Forrestel, E.J., Earles, J.M., 2020. Digitally deconstructing leaves in 3D using X-ray microcomputed tomography and machine learning. Applications in plant sciences 8, e11380.

Théroux-Rancourt, G., Roddy, A.B., Earles, J.M., Gilbert, M.E., Zwieniecki, M.A., Boyce, C.K., Tholen, D., McElrone, A.J., Simonin, K.A., Brodersen, C.R., 2021. Maximum $CO_2$ diffusion inside leaves is limited by the scaling of cell size and genome size. Proceedings of the Royal Society B 288, 20203145.

Torres-Ruiz, J.M., Jansen, S., Choat, B., McElrone, A.J., Cochard, H., Brodribb, T.J., Badel, E., Burlett, R., Bouche, P.S., Brodersen, C.R., 2015. Direct X-ray microtomography observation confirms the induction of embolism upon xylem cutting under tension. Plant Physiology 167, 40–43.

Tracy, S.R., Roberts, J.A., Black, C.R., McNeill, A., Davidson, R., Mooney, S.J., 2010. The X-factor: visualizing undisturbed root architecture in soils using X-ray computed tomography. Journal of experimental botany 61, 311–313.

von Chamier, L., Laine, R.F., Jukkala, J., Spahn, C., Krentzel, D., Nehme, E., Lerche, M., Hernández-Pérez, S., Mattila, P.K., Karinou, E., Holden, S., Solak, A.C., Krull, A., Buchholz, T.-O., Jones, M.L., Royer, L.A., Leterrier, C., Shechtman, Y., Jug, F., Heilemann, M., Jacquemet, G., Henriques, R., 2021. Democratising deep learning for microscopy with ZeroCostDL4Mic. Nat Commun 12, 2276. https://doi.org/10.1038/s41467-021-22518-0

Waskom, M.L., 2021. Seaborn: statistical data visualization. Journal of Open Source Software 6, 3021.

Xiao, Z., Stait-Gardner, T., Willis, S.A., Price, W.S., Moroni, F.J., Pagay, V., Tyerman, S.D., Schmidtke, L.M., Rogiers, S.Y., 2021. 3D visualisation of voids in grapevine flowers and berries using X-ray micro computed tomography. Australian Journal of Grape and Wine Research 27, 141–148. https://doi.org/10.1111/ajgw.12480

Yudina, A., Kuzyakov, Y., 2019. Saving the face of soil aggregates. Global Change Biology 25, 3574–3577. https://doi.org/10.1111/gcb.14779

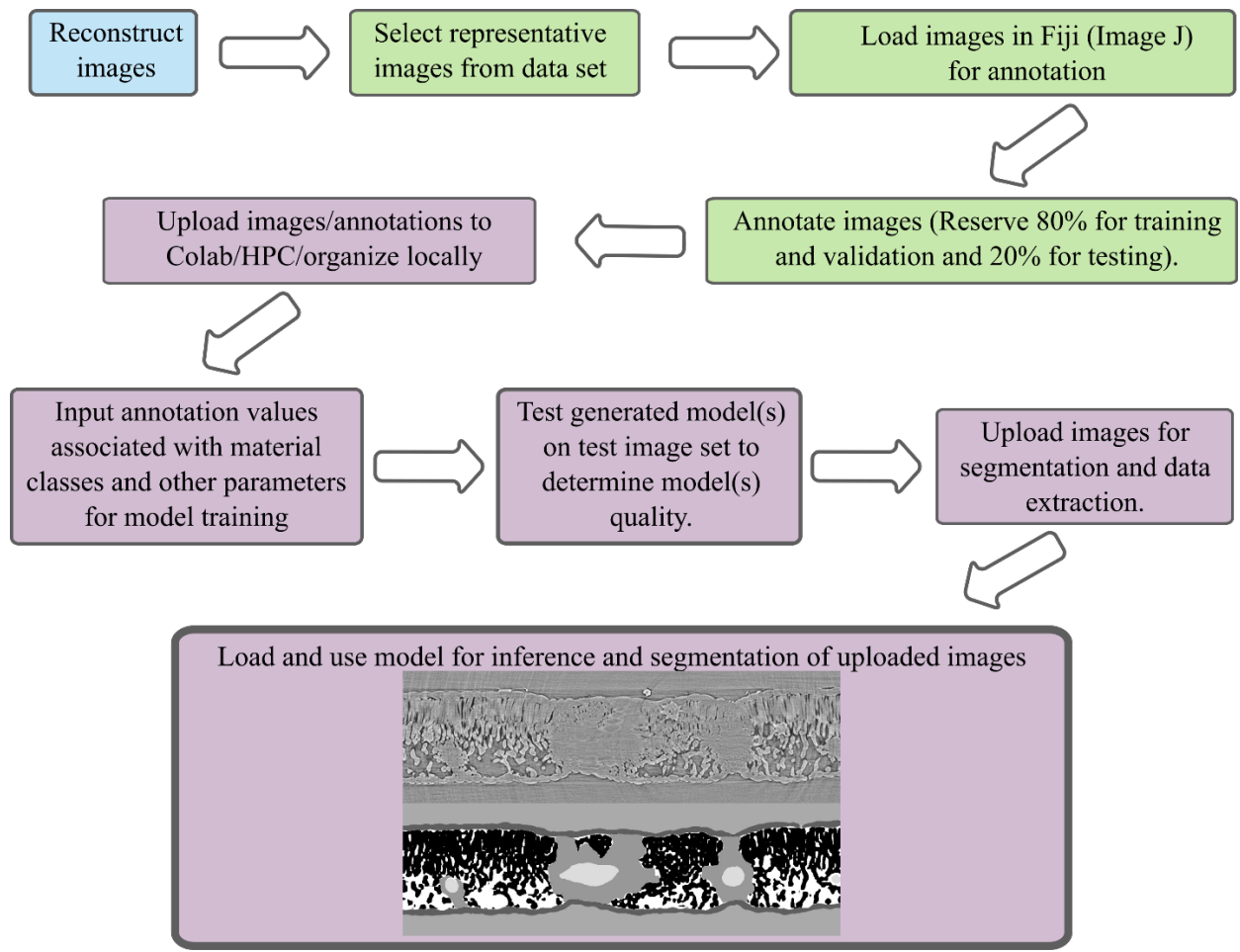

Figure 1: A schematic of the segmentation workflow from image reconstruction, image annotation, model training, model use, and data extraction. Blue indicates a process that is done at the instrumentation site, green is a process done on local computers using a subset of the data in ImageJ or CVAT, purple indicates a process done in Google's Colaboratory, on a high performance computing cluster, or locally.

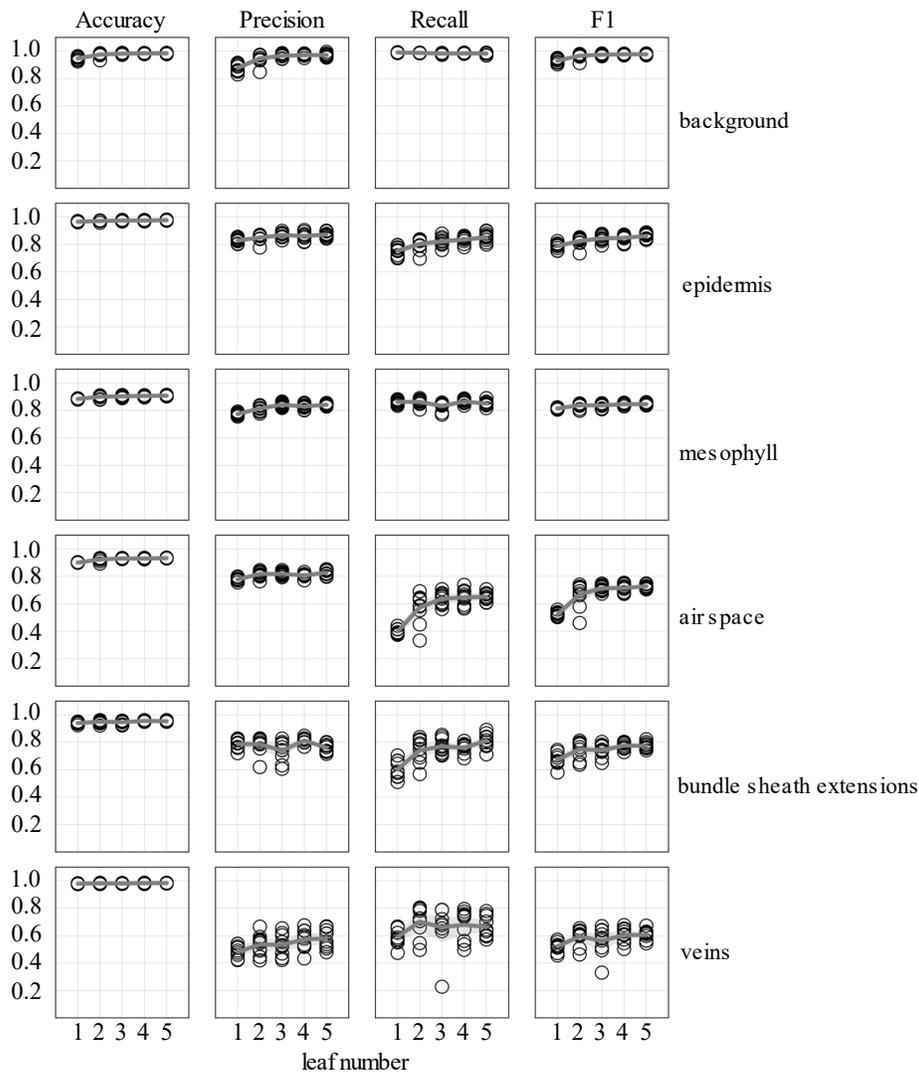

Figure 2: Accuracy, precision, recall, and F1 scores as a function of uniquely annotated leaf number (5 annotated images per leaf) used to predict tissue classes in X-ray µCT images from an independent leaf on which the models were not trained or validated. Circles represent unique predictions from 10 uniquely generated models per leaf number; dark gray lines represent the mean value of the 10 models while thick light gray lines represent the 95% confidence interval of the values.

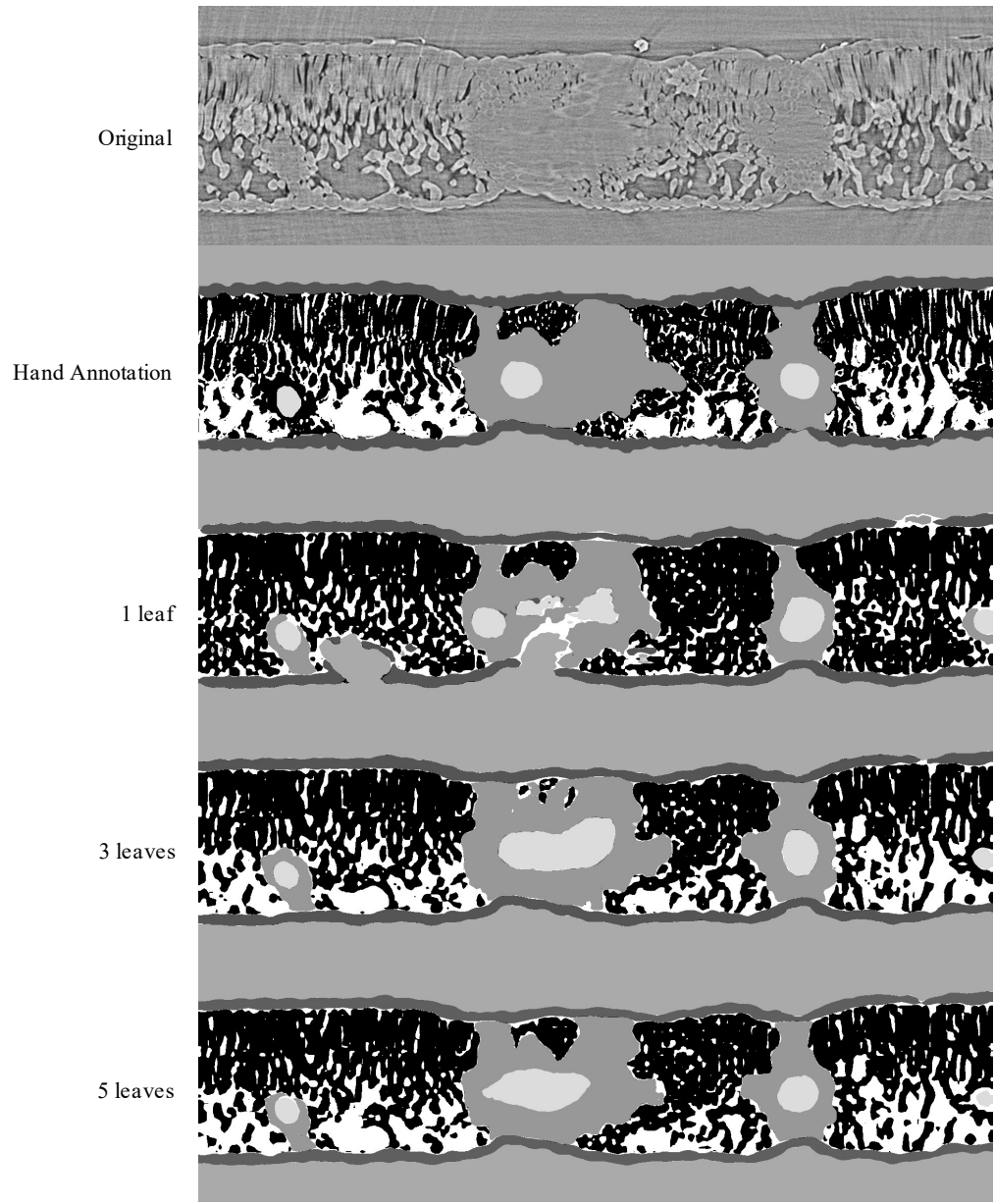

Figure 3: Visual representation of the model outputs for a single walnut leaf image; top image is a X-ray CT scan taken from a leaf that was not used for training or validation of the applied models; next is the hand annotated image of the scan followed by the outputs from the best performing model trained on 1, 3, and 5 leaves, respectively. For the walnut leaf segmentations

the background is light gray, the epidermis is dark gray, the mesophyll is black, the air space is white, the bundle sheath extensions are middle gray, and the veins are lightest gray.

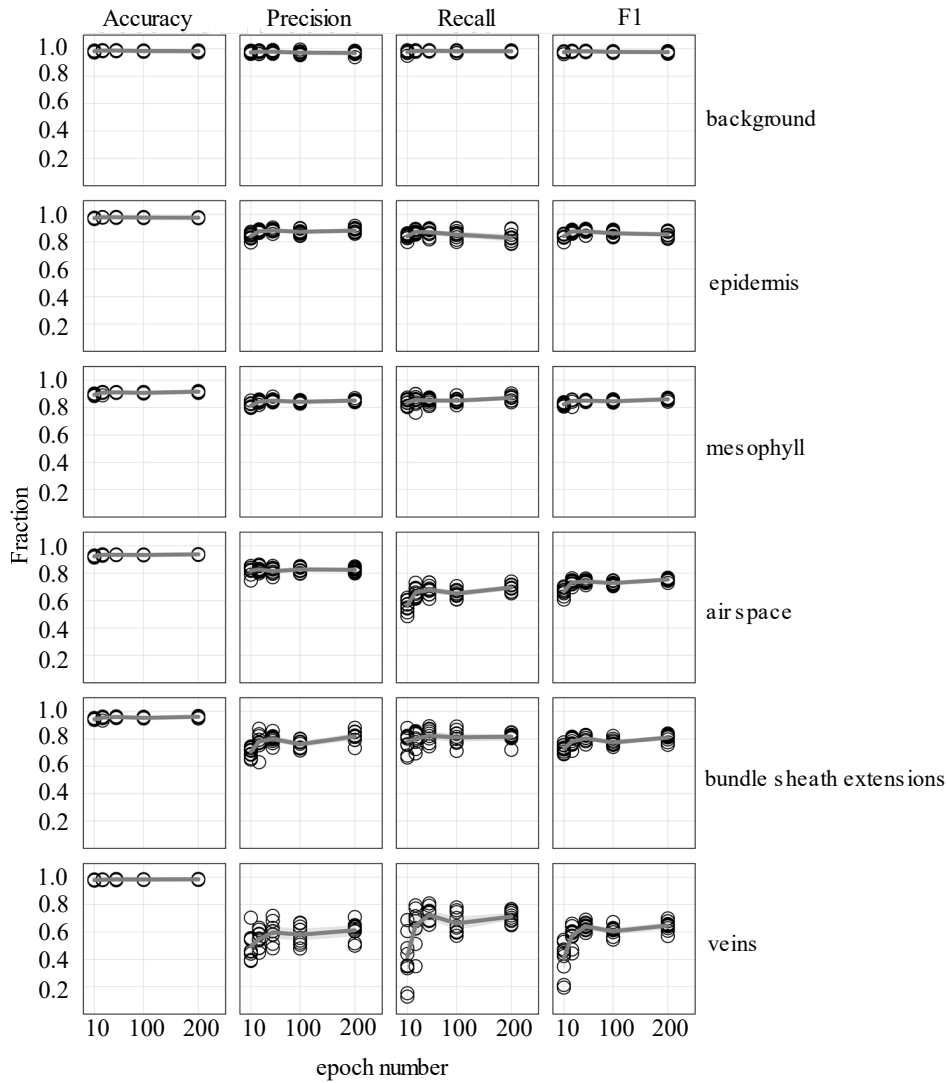

Figure 4: Accuracy, precision, recall, and F1 scores as a function increasing epoch number for models trained on annotated images from 5 leaves (5 annotated images per leaf) used to predict tissue classes from X-ray µCT images from an independent leaf on which the models were not trained or validated. Circles represent unique predictions from 10 uniquely generated models per epoch number; dark gray lines represent the mean value of the 10 models for each tissue type while thick light gray lines represent the 95% confidence interval of the values.

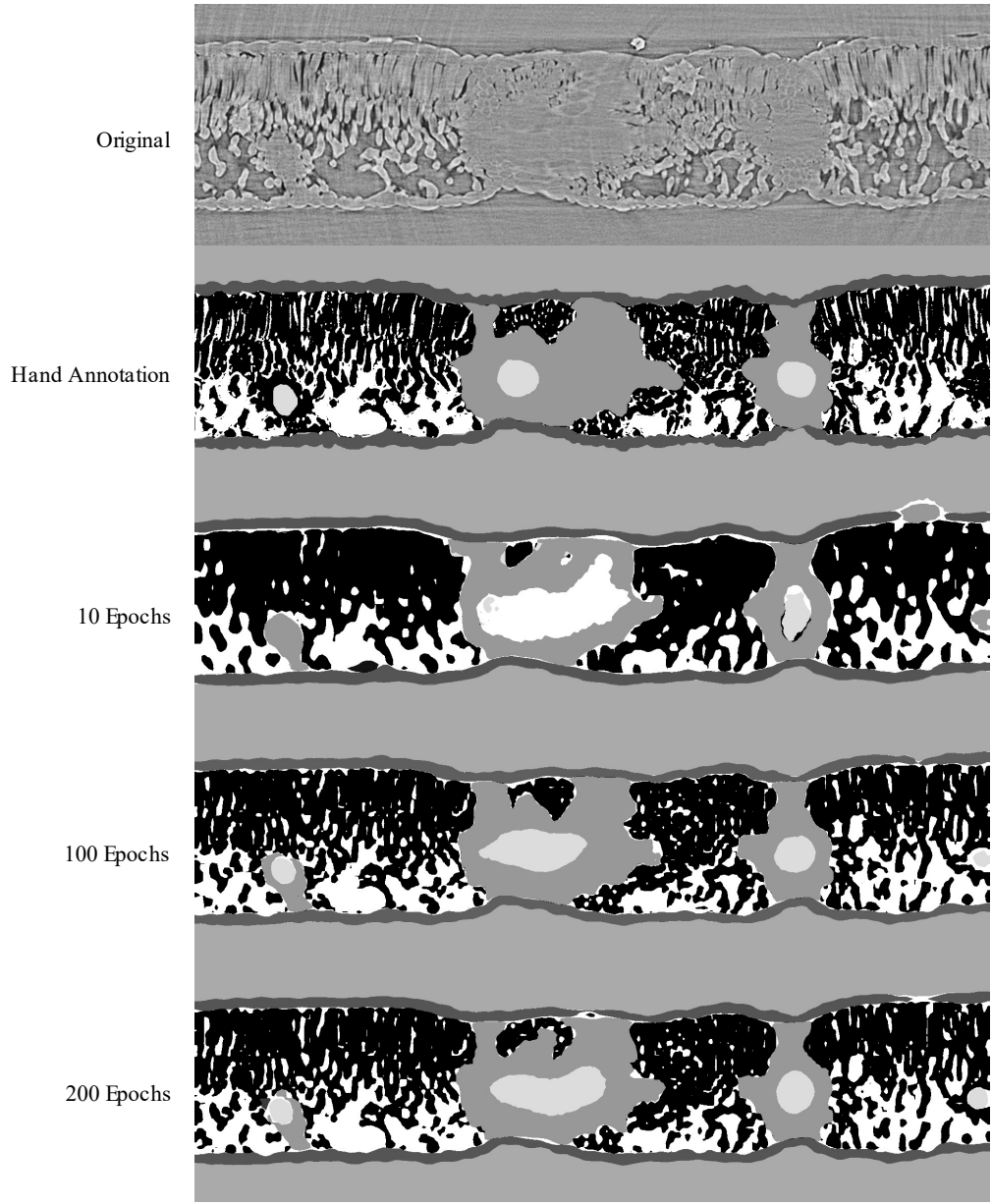

Figure 5: Visual representation of the model outputs for a single walnut leaf image cross-section; top image is a X-ray CT scan taken from a leaf that was not used for training or validation of the applied models; next is the hand annotated image of the scan followed by the outputs from the best performing model trained for 10, 100, and 200 epochs respectively. For the walnut leaf

segmentation, the background is light gray, the epidermis is dark gray, the mesophyll is black, the air space is white, the bundle sheath extensions are middle gray, and the veins are lightest gray.

| Almond Bud | Soil Aggregate |

Original

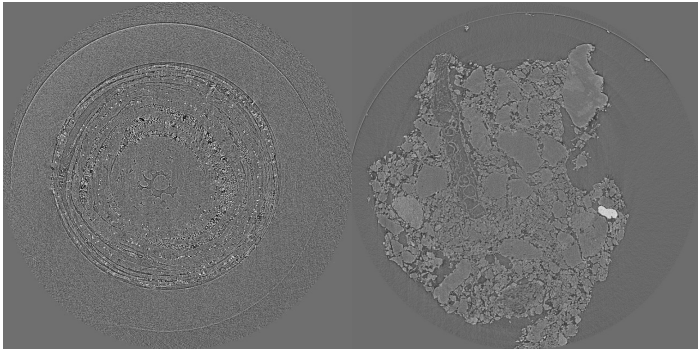

Hand

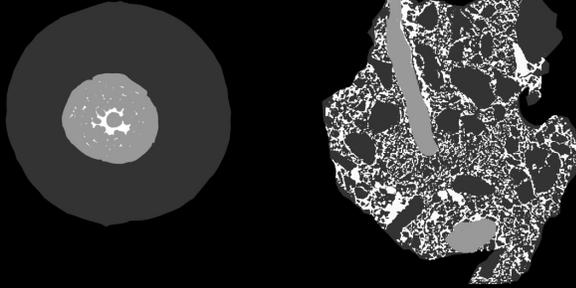

0.5 Scale

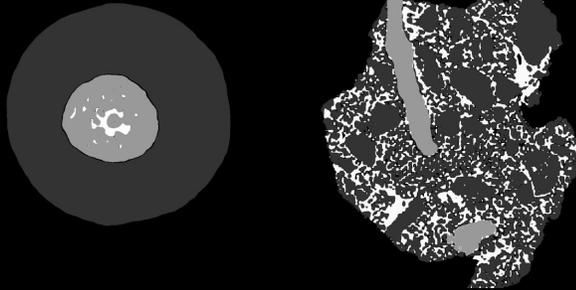

0.85 Scale

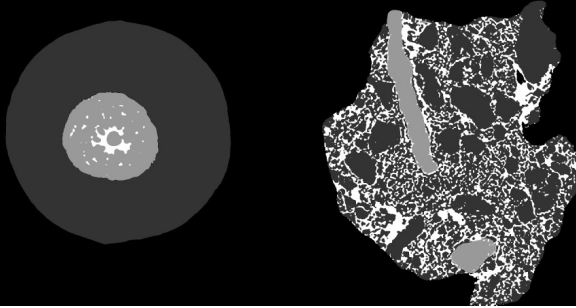

Figure 6: Visual representation of the model outputs for an almond flower bud and a soil aggregate; top images are X-ray CT scans of the flower bud and soil aggregate; next are the hand annotated images of the scans followed by the outputs from the best performing models trained at 0.5 and 0.85 scale, respectively. For the almond flower bud, background is black, bud scales are dark gray, flower tissue is light gray, and air spaces are white. For the soil aggregate, background is black, mineral solids are dark gray, pore spaces are white, and organic matter is light gray.

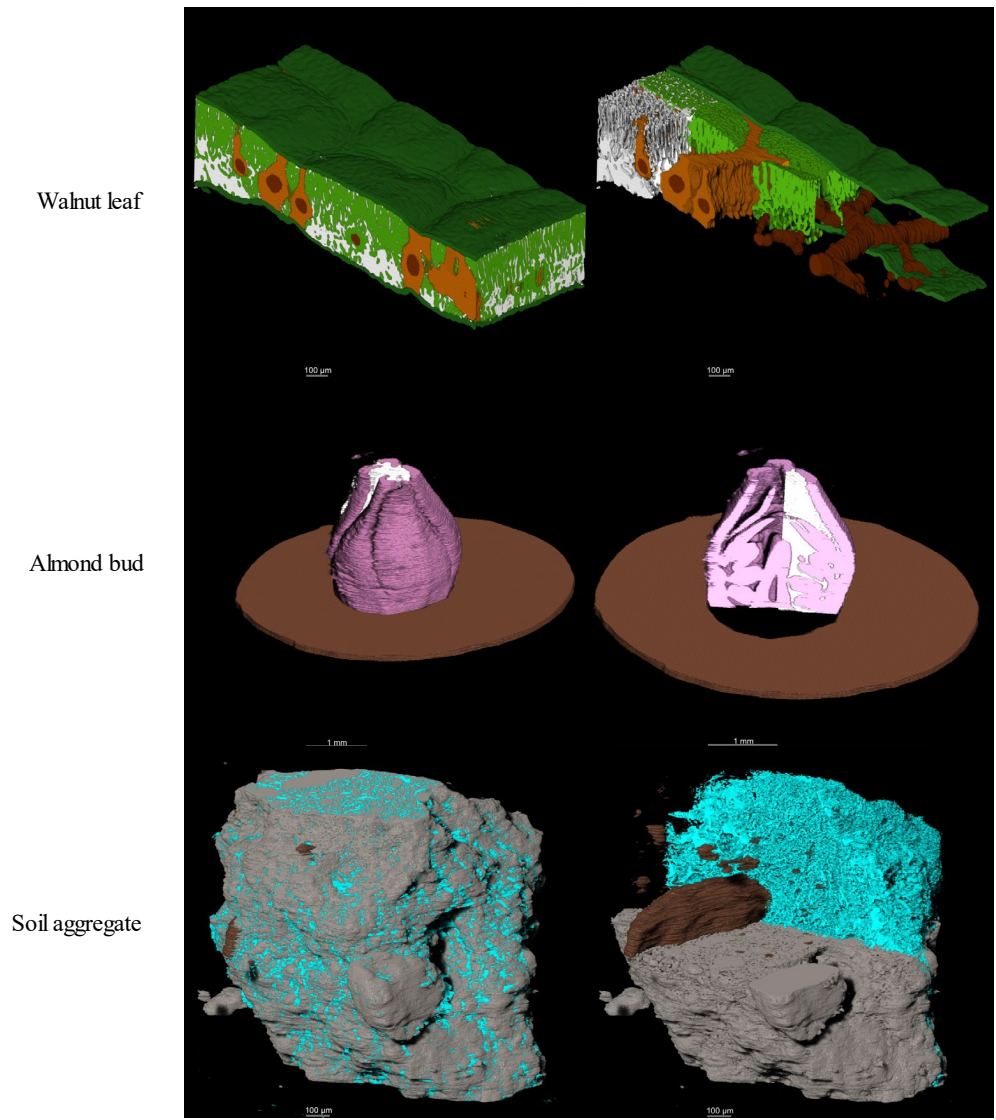

Figure 7: 3D visual representation of the stacked model outputs for a walnut leaf, a almond flower bud, and a soil aggregate; top images are X-ray CT scans of the bud and soil aggregate; next are the hand. In the walnut leaf images, the background is black, the epidermis is dark green, the mesophyll is light green, the air space is white, the bundle sheath extensions burnt orange, and the veins are brown. In the almond bud images, background is black, bud scales are brown, flower tissue is pink, and air spaces are white. For the soil aggregate, background is black, mineral solids are dark gray, pore spaces are blue, and organic matter is brown.

SI Table 2: Tabular representation of the sample types, scale, unique scan number, training/validation image number, and testing image number used to generate various models use in the paper.

| Sample Type | Scale | Unique Scan # | Training/Validation Image # | Testing Image # |
|---|---|---|---|---|
| Walnut Leaf | 1 | 1 | 5 | 5 |
| | 1 | 2 | 10 | 5 |
| | 1 | 3 | 15 | 5 |
| | 1 | 4 | 20 | 5 |
| | 1 | 5 | 25 | 5 |
| Almond Bud | 0.5 | 1 | 7 | 7 |
| | 0.85 | 1 | 7 | 7 |
| Soil Aggregate | 0.5 | 1 | 11 | 5 |
| | 0.85 | 1 | 11 | 5 |

SI Table 2: Tabular representation of a Pytorch implementation of a Fully Convolutional Network with a Res-net 101 backbone with an input image size of 1000x500 pixels with 6 material classes.

| Layer (type:depth-idx) | Output Shape | Param # |
|---|---|---|
| FCN | [1, 6, 1000, 500] | -- |
| ├─IntermediateLayerGetter: | [1, 2048, 125, 63] | -- |
| │    └─Conv2d: | [1, 64, 500, 250] | 9,408 |
| │    └─BatchNorm2d: | [1, 64, 500, 250] | 128 |
| │    └─ReLU: | [1, 64, 500, 250] | -- |
| │    └─MaxPool2d: | [1, 64, 250, 125] | -- |
| │    └─Sequential: 1 | [1, 256, 250, 125] | -- |
| │    │    └─Bottleneck: 0 | [1, 256, 250, 125] | 75,008 |
| │    │    └─Bottleneck: 1 | [1, 256, 250, 125] | 70,400 |
| │    │    └─Bottleneck: 2 | [1, 256, 250, 125] | 70,400 |
| │    └─Sequential: 2 | [1, 512, 125, 63] | -- |
| │    │    └─Bottleneck: 0 | [1, 512, 125, 63] | 379,392 |
| │    │    └─Bottleneck: 1 | [1, 512, 125, 63] | 280,064 |
| │    │    └─Bottleneck: 2 | [1, 512, 125, 63] | 280,064 |
| │    │    └─Bottleneck: 3 | [1, 512, 125, 63] | 280,064 |
| │    └─Sequential: 3 | [1, 1024, 125, 63] | -- |
| │    │    └─Bottleneck: 0 | [1, 1024, 125, 63] | 1,512,448 |
| │    │    └─Bottleneck: 1 | [1, 1024, 125, 63] | 1,117,184 |
| │    │    └─Bottleneck: 2 | [1, 1024, 125, 63] | 1,117,184 |
| │    │    └─Bottleneck: 3 | [1, 1024, 125, 63] | 1,117,184 |
| │    │    └─Bottleneck: 4 | [1, 1024, 125, 63] | 1,117,184 |
| │    │    └─Bottleneck: 5 | [1, 1024, 125, 63] | 1,117,184 |
| │    │    └─Bottleneck: 6 | [1, 1024, 125, 63] | 1,117,184 |

```
|   |   └─Bottleneck: 7           [1, 1024, 125, 63]     1,117,184
|   |   └─Bottleneck: 8           [1, 1024, 125, 63]     1,117,184
|   |   └─Bottleneck: 9           [1, 1024, 125, 63]     1,117,184
|   |   └─Bottleneck: 10          [1, 1024, 125, 63]     1,117,184
|   |   └─Bottleneck: 11          [1, 1024, 125, 63]     1,117,184
|   |   └─Bottleneck: 12          [1, 1024, 125, 63]     1,117,184
|   |   └─Bottleneck: 13          [1, 1024, 125, 63]     1,117,184
|   |   └─Bottleneck: 14          [1, 1024, 125, 63]     1,117,184
|   |   └─Bottleneck: 15          [1, 1024, 125, 63]     1,117,184
|   |   └─Bottleneck: 16          [1, 1024, 125, 63]     1,117,184
|   |   └─Bottleneck: 17          [1, 1024, 125, 63]     1,117,184
|   |   └─Bottleneck: 18          [1, 1024, 125, 63]     1,117,184
|   |   └─Bottleneck: 19          [1, 1024, 125, 63]     1,117,184
|   |   └─Bottleneck: 20          [1, 1024, 125, 63]     1,117,184
|   |   └─Bottleneck: 21          [1, 1024, 125, 63]     1,117,184
|   |   └─Bottleneck: 22          [1, 1024, 125, 63]     1,117,184
|   └─Sequential: 4               [1, 2048, 125, 63]     --
|   |   └─Bottleneck: 0           [1, 2048, 125, 63]     6,039,552
|   |   └─Bottleneck: 1           [1, 2048, 125, 63]     4,462,592
|   |   └─Bottleneck: 2           [1, 2048, 125, 63]     4,462,592
├─FCNHead:                        [1, 6, 125, 63]        --
|   └─Conv2d: 0                   [1, 512, 125, 63]      9,437,184
|   └─BatchNorm2d: 1              [1, 512, 125, 63]      1,024
|   └─ReLU: 2                     [1, 512, 125, 63]      --
|   └─Dropout: 3                  [1, 512, 125, 63]      --
|   └─Conv2d: 4                   [1, 6, 125, 63]        3,078
=================================================================
=========================
Total params: 51,941,446
```

Trainable params: 51,941,446

Non-trainable params: 0

Total mult-adds (G): 415.05

================================================================================

Input size (MB): 6.00

Forward/backward pass size (MB): 7395.96

Params size (MB): 207.77

Estimated Total Size (MB): 7609.73

================================================================================

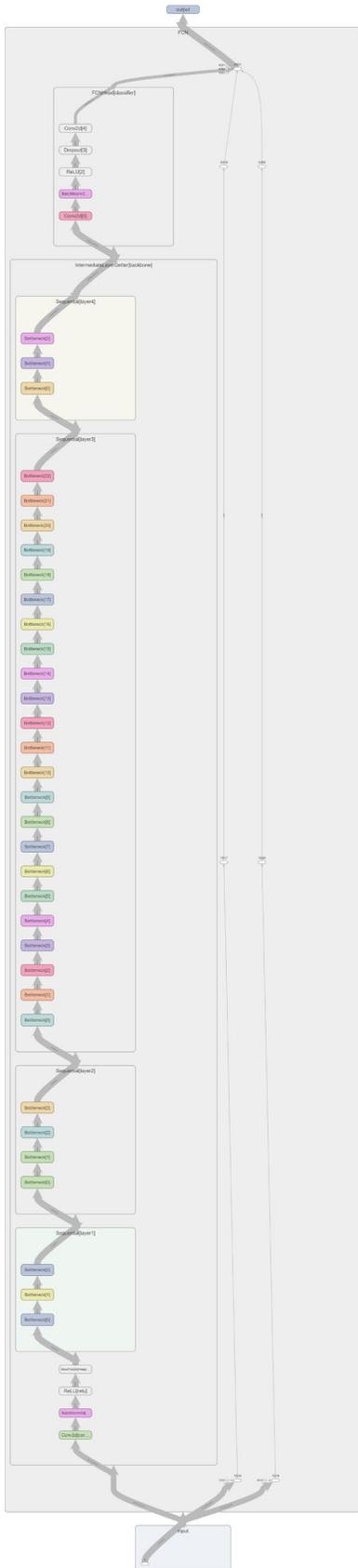

SI Figure 1: Generalized structure of a Pytorch implementation of a Fully Convolutional Network with a Res-net 101 backbone. (SVG image, please zoom in for details).

SI Figure 2: Detailed summary of the individual processes occurring in each layer of a Pytorch implementation of a Fully Convolutional Network with a Res-net 101 backbone.

```
FCN(
  (backbone): IntermediateLayerGetter(
    (conv1): Conv2d(3, 64, kernel_size=(7, 7), stride=(2, 2), padding=(3, 3), bias=False)
    (bn1): BatchNorm2d(64, eps=1e-05, momentum=0.1, affine=True, track_running_stats=True)
    (relu): ReLU(inplace=True)
    (maxpool): MaxPool2d(kernel_size=3, stride=2, padding=1, dilation=1, ceil_mode=False)
    (layer1): Sequential(
      (0): Bottleneck(
        (conv1): Conv2d(64, 64, kernel_size=(1, 1), stride=(1, 1), bias=False)
        (bn1): BatchNorm2d(64, eps=1e-05, momentum=0.1, affine=True, track_running_stats=True)
        (conv2): Conv2d(64, 64, kernel_size=(3, 3), stride=(1, 1), padding=(1, 1), bias=False)
        (bn2): BatchNorm2d(64, eps=1e-05, momentum=0.1, affine=True, track_running_stats=True)
        (conv3): Conv2d(64, 256, kernel_size=(1, 1), stride=(1, 1), bias=False)
        (bn3): BatchNorm2d(256, eps=1e-05, momentum=0.1, affine=True, track_running_stats=True)
        (relu): ReLU(inplace=True)
        (downsample): Sequential(
          (0): Conv2d(64, 256, kernel_size=(1, 1), stride=(1, 1), bias=False)
          (1): BatchNorm2d(256, eps=1e-05, momentum=0.1, affine=True, track_running_stats=True)
        )
      )
      (1): Bottleneck(
        (conv1): Conv2d(256, 64, kernel_size=(1, 1), stride=(1, 1), bias=False)
```

```
      (bn1): BatchNorm2d(64, eps=1e-05, momentum=0.1, affine=True, track_running_stats=True)
      (conv2): Conv2d(64, 64, kernel_size=(3, 3), stride=(1, 1), padding=(1, 1), bias=False)
      (bn2): BatchNorm2d(64, eps=1e-05, momentum=0.1, affine=True, track_running_stats=True)
      (conv3): Conv2d(64, 256, kernel_size=(1, 1), stride=(1, 1), bias=False)
      (bn3): BatchNorm2d(256, eps=1e-05, momentum=0.1, affine=True, track_running_stats=True)
      (relu): ReLU(inplace=True)
    )
    (2): Bottleneck(
      (conv1): Conv2d(256, 64, kernel_size=(1, 1), stride=(1, 1), bias=False)
      (bn1): BatchNorm2d(64, eps=1e-05, momentum=0.1, affine=True, track_running_stats=True)
      (conv2): Conv2d(64, 64, kernel_size=(3, 3), stride=(1, 1), padding=(1, 1), bias=False)
      (bn2): BatchNorm2d(64, eps=1e-05, momentum=0.1, affine=True, track_running_stats=True)
      (conv3): Conv2d(64, 256, kernel_size=(1, 1), stride=(1, 1), bias=False)
      (bn3): BatchNorm2d(256, eps=1e-05, momentum=0.1, affine=True, track_running_stats=True)
      (relu): ReLU(inplace=True)
    )
  )
  (layer2): Sequential(
    (0): Bottleneck(
      (conv1): Conv2d(256, 128, kernel_size=(1, 1), stride=(1, 1), bias=False)
      (bn1): BatchNorm2d(128, eps=1e-05, momentum=0.1, affine=True, track_running_stats=True)
      (conv2): Conv2d(128, 128, kernel_size=(3, 3), stride=(2, 2), padding=(1, 1), bias=False)
      (bn2): BatchNorm2d(128, eps=1e-05, momentum=0.1, affine=True, track_running_stats=True)
```

(conv3): Conv2d(128, 512, kernel_size=(1, 1), stride=(1, 1), bias=False)

        (bn3): BatchNorm2d(512, eps=1e-05, momentum=0.1, affine=True, track_running_stats=True)

        (relu): ReLU(inplace=True)

        (downsample): Sequential(

          (0): Conv2d(256, 512, kernel_size=(1, 1), stride=(2, 2), bias=False)

          (1): BatchNorm2d(512, eps=1e-05, momentum=0.1, affine=True, track_running_stats=True)

        )

      )

      (1): Bottleneck(

        (conv1): Conv2d(512, 128, kernel_size=(1, 1), stride=(1, 1), bias=False)

        (bn1): BatchNorm2d(128, eps=1e-05, momentum=0.1, affine=True, track_running_stats=True)

        (conv2): Conv2d(128, 128, kernel_size=(3, 3), stride=(1, 1), padding=(1, 1), bias=False)

        (bn2): BatchNorm2d(128, eps=1e-05, momentum=0.1, affine=True, track_running_stats=True)

        (conv3): Conv2d(128, 512, kernel_size=(1, 1), stride=(1, 1), bias=False)

        (bn3): BatchNorm2d(512, eps=1e-05, momentum=0.1, affine=True, track_running_stats=True)

        (relu): ReLU(inplace=True)

      )

      (2): Bottleneck(

        (conv1): Conv2d(512, 128, kernel_size=(1, 1), stride=(1, 1), bias=False)

        (bn1): BatchNorm2d(128, eps=1e-05, momentum=0.1, affine=True, track_running_stats=True)

        (conv2): Conv2d(128, 128, kernel_size=(3, 3), stride=(1, 1), padding=(1, 1), bias=False)

        (bn2): BatchNorm2d(128, eps=1e-05, momentum=0.1, affine=True, track_running_stats=True)

        (conv3): Conv2d(128, 512, kernel_size=(1, 1), stride=(1, 1), bias=False)

(bn3): BatchNorm2d(512, eps=1e-05, momentum=0.1, affine=True, track_running_stats=True)

      (relu): ReLU(inplace=True)

    )

    (3): Bottleneck(

      (conv1): Conv2d(512, 128, kernel_size=(1, 1), stride=(1, 1), bias=False)

      (bn1): BatchNorm2d(128, eps=1e-05, momentum=0.1, affine=True, track_running_stats=True)

      (conv2): Conv2d(128, 128, kernel_size=(3, 3), stride=(1, 1), padding=(1, 1), bias=False)

      (bn2): BatchNorm2d(128, eps=1e-05, momentum=0.1, affine=True, track_running_stats=True)

      (conv3): Conv2d(128, 512, kernel_size=(1, 1), stride=(1, 1), bias=False)

      (bn3): BatchNorm2d(512, eps=1e-05, momentum=0.1, affine=True, track_running_stats=True)

      (relu): ReLU(inplace=True)

    )

  )

  (layer3): Sequential(

    (0): Bottleneck(

      (conv1): Conv2d(512, 256, kernel_size=(1, 1), stride=(1, 1), bias=False)

      (bn1): BatchNorm2d(256, eps=1e-05, momentum=0.1, affine=True, track_running_stats=True)

      (conv2): Conv2d(256, 256, kernel_size=(3, 3), stride=(1, 1), padding=(1, 1), bias=False)

      (bn2): BatchNorm2d(256, eps=1e-05, momentum=0.1, affine=True, track_running_stats=True)

      (conv3): Conv2d(256, 1024, kernel_size=(1, 1), stride=(1, 1), bias=False)

      (bn3): BatchNorm2d(1024, eps=1e-05, momentum=0.1, affine=True, track_running_stats=True)

      (relu): ReLU(inplace=True)

      (downsample): Sequential(

        (0): Conv2d(512, 1024, kernel_size=(1, 1), stride=(1, 1), bias=False)

```
      (1): BatchNorm2d(1024, eps=1e-05, momentum=0.1, affine=True, track_running_stats=True)
    )
  )
  (1): Bottleneck(
    (conv1): Conv2d(1024, 256, kernel_size=(1, 1), stride=(1, 1), bias=False)
    (bn1): BatchNorm2d(256, eps=1e-05, momentum=0.1, affine=True, track_running_stats=True)
    (conv2): Conv2d(256, 256, kernel_size=(3, 3), stride=(1, 1), padding=(2, 2), dilation=(2, 2), bias=False)
    (bn2): BatchNorm2d(256, eps=1e-05, momentum=0.1, affine=True, track_running_stats=True)
    (conv3): Conv2d(256, 1024, kernel_size=(1, 1), stride=(1, 1), bias=False)
    (bn3): BatchNorm2d(1024, eps=1e-05, momentum=0.1, affine=True, track_running_stats=True)
    (relu): ReLU(inplace=True)
  )
  (2): Bottleneck(
    (conv1): Conv2d(1024, 256, kernel_size=(1, 1), stride=(1, 1), bias=False)
    (bn1): BatchNorm2d(256, eps=1e-05, momentum=0.1, affine=True, track_running_stats=True)
    (conv2): Conv2d(256, 256, kernel_size=(3, 3), stride=(1, 1), padding=(2, 2), dilation=(2, 2), bias=False)
    (bn2): BatchNorm2d(256, eps=1e-05, momentum=0.1, affine=True, track_running_stats=True)
    (conv3): Conv2d(256, 1024, kernel_size=(1, 1), stride=(1, 1), bias=False)
    (bn3): BatchNorm2d(1024, eps=1e-05, momentum=0.1, affine=True, track_running_stats=True)
    (relu): ReLU(inplace=True)
  )
  (3): Bottleneck(
```

```
      (conv1): Conv2d(1024, 256, kernel_size=(1, 1), stride=(1, 1), bias=False)
      (bn1): BatchNorm2d(256, eps=1e-05, momentum=0.1, affine=True, track_running_stats=True)
      (conv2): Conv2d(256, 256, kernel_size=(3, 3), stride=(1, 1), padding=(2, 2), dilation=(2, 2), bias=False)
      (bn2): BatchNorm2d(256, eps=1e-05, momentum=0.1, affine=True, track_running_stats=True)
      (conv3): Conv2d(256, 1024, kernel_size=(1, 1), stride=(1, 1), bias=False)
      (bn3): BatchNorm2d(1024, eps=1e-05, momentum=0.1, affine=True, track_running_stats=True)
      (relu): ReLU(inplace=True)
    )
    (4): Bottleneck(
      (conv1): Conv2d(1024, 256, kernel_size=(1, 1), stride=(1, 1), bias=False)
      (bn1): BatchNorm2d(256, eps=1e-05, momentum=0.1, affine=True, track_running_stats=True)
      (conv2): Conv2d(256, 256, kernel_size=(3, 3), stride=(1, 1), padding=(2, 2), dilation=(2, 2), bias=False)
      (bn2): BatchNorm2d(256, eps=1e-05, momentum=0.1, affine=True, track_running_stats=True)
      (conv3): Conv2d(256, 1024, kernel_size=(1, 1), stride=(1, 1), bias=False)
      (bn3): BatchNorm2d(1024, eps=1e-05, momentum=0.1, affine=True, track_running_stats=True)
      (relu): ReLU(inplace=True)
    )
    (5): Bottleneck(
      (conv1): Conv2d(1024, 256, kernel_size=(1, 1), stride=(1, 1), bias=False)
      (bn1): BatchNorm2d(256, eps=1e-05, momentum=0.1, affine=True, track_running_stats=True)
      (conv2): Conv2d(256, 256, kernel_size=(3, 3), stride=(1, 1), padding=(2, 2), dilation=(2, 2), bias=False)
```

(bn2): BatchNorm2d(256, eps=1e-05, momentum=0.1, affine=True, track_running_stats=True)

      (conv3): Conv2d(256, 1024, kernel_size=(1, 1), stride=(1, 1), bias=False)

      (bn3): BatchNorm2d(1024, eps=1e-05, momentum=0.1, affine=True, track_running_stats=True)

      (relu): ReLU(inplace=True)

    )

    (6): Bottleneck(

      (conv1): Conv2d(1024, 256, kernel_size=(1, 1), stride=(1, 1), bias=False)

      (bn1): BatchNorm2d(256, eps=1e-05, momentum=0.1, affine=True, track_running_stats=True)

      (conv2): Conv2d(256, 256, kernel_size=(3, 3), stride=(1, 1), padding=(2, 2), dilation=(2, 2), bias=False)

      (bn2): BatchNorm2d(256, eps=1e-05, momentum=0.1, affine=True, track_running_stats=True)

      (conv3): Conv2d(256, 1024, kernel_size=(1, 1), stride=(1, 1), bias=False)

      (bn3): BatchNorm2d(1024, eps=1e-05, momentum=0.1, affine=True, track_running_stats=True)

      (relu): ReLU(inplace=True)

    )

    (7): Bottleneck(

      (conv1): Conv2d(1024, 256, kernel_size=(1, 1), stride=(1, 1), bias=False)

      (bn1): BatchNorm2d(256, eps=1e-05, momentum=0.1, affine=True, track_running_stats=True)

      (conv2): Conv2d(256, 256, kernel_size=(3, 3), stride=(1, 1), padding=(2, 2), dilation=(2, 2), bias=False)

      (bn2): BatchNorm2d(256, eps=1e-05, momentum=0.1, affine=True, track_running_stats=True)

      (conv3): Conv2d(256, 1024, kernel_size=(1, 1), stride=(1, 1), bias=False)

      (bn3): BatchNorm2d(1024, eps=1e-05, momentum=0.1, affine=True, track_running_stats=True)

      (relu): ReLU(inplace=True)

```
    )
    (8): Bottleneck(
      (conv1): Conv2d(1024, 256, kernel_size=(1, 1), stride=(1, 1), bias=False)
      (bn1): BatchNorm2d(256, eps=1e-05, momentum=0.1, affine=True, track_running_stats=True)
      (conv2): Conv2d(256, 256, kernel_size=(3, 3), stride=(1, 1), padding=(2, 2), dilation=(2, 2), bias=False)
      (bn2): BatchNorm2d(256, eps=1e-05, momentum=0.1, affine=True, track_running_stats=True)
      (conv3): Conv2d(256, 1024, kernel_size=(1, 1), stride=(1, 1), bias=False)
      (bn3): BatchNorm2d(1024, eps=1e-05, momentum=0.1, affine=True, track_running_stats=True)
      (relu): ReLU(inplace=True)
    )
    (9): Bottleneck(
      (conv1): Conv2d(1024, 256, kernel_size=(1, 1), stride=(1, 1), bias=False)
      (bn1): BatchNorm2d(256, eps=1e-05, momentum=0.1, affine=True, track_running_stats=True)
      (conv2): Conv2d(256, 256, kernel_size=(3, 3), stride=(1, 1), padding=(2, 2), dilation=(2, 2), bias=False)
      (bn2): BatchNorm2d(256, eps=1e-05, momentum=0.1, affine=True, track_running_stats=True)
      (conv3): Conv2d(256, 1024, kernel_size=(1, 1), stride=(1, 1), bias=False)
      (bn3): BatchNorm2d(1024, eps=1e-05, momentum=0.1, affine=True, track_running_stats=True)
      (relu): ReLU(inplace=True)
    )
    (10): Bottleneck(
      (conv1): Conv2d(1024, 256, kernel_size=(1, 1), stride=(1, 1), bias=False)
      (bn1): BatchNorm2d(256, eps=1e-05, momentum=0.1, affine=True, track_running_stats=True)
```

(conv2): Conv2d(256, 256, kernel_size=(3, 3), stride=(1, 1), padding=(2, 2), dilation=(2, 2), bias=False)

      (bn2): BatchNorm2d(256, eps=1e-05, momentum=0.1, affine=True, track_running_stats=True)

      (conv3): Conv2d(256, 1024, kernel_size=(1, 1), stride=(1, 1), bias=False)

      (bn3): BatchNorm2d(1024, eps=1e-05, momentum=0.1, affine=True, track_running_stats=True)

      (relu): ReLU(inplace=True)

    )

    (11): Bottleneck(

      (conv1): Conv2d(1024, 256, kernel_size=(1, 1), stride=(1, 1), bias=False)

      (bn1): BatchNorm2d(256, eps=1e-05, momentum=0.1, affine=True, track_running_stats=True)

      (conv2): Conv2d(256, 256, kernel_size=(3, 3), stride=(1, 1), padding=(2, 2), dilation=(2, 2), bias=False)

      (bn2): BatchNorm2d(256, eps=1e-05, momentum=0.1, affine=True, track_running_stats=True)

      (conv3): Conv2d(256, 1024, kernel_size=(1, 1), stride=(1, 1), bias=False)

      (bn3): BatchNorm2d(1024, eps=1e-05, momentum=0.1, affine=True, track_running_stats=True)

      (relu): ReLU(inplace=True)

    )

    (12): Bottleneck(

      (conv1): Conv2d(1024, 256, kernel_size=(1, 1), stride=(1, 1), bias=False)

      (bn1): BatchNorm2d(256, eps=1e-05, momentum=0.1, affine=True, track_running_stats=True)

      (conv2): Conv2d(256, 256, kernel_size=(3, 3), stride=(1, 1), padding=(2, 2), dilation=(2, 2), bias=False)

      (bn2): BatchNorm2d(256, eps=1e-05, momentum=0.1, affine=True, track_running_stats=True)

      (conv3): Conv2d(256, 1024, kernel_size=(1, 1), stride=(1, 1), bias=False)

```
    (bn3): BatchNorm2d(1024, eps=1e-05, momentum=0.1, affine=True, track_running_stats=True)
    (relu): ReLU(inplace=True)
  )
  (13): Bottleneck(
    (conv1): Conv2d(1024, 256, kernel_size=(1, 1), stride=(1, 1), bias=False)
    (bn1): BatchNorm2d(256, eps=1e-05, momentum=0.1, affine=True, track_running_stats=True)
    (conv2): Conv2d(256, 256, kernel_size=(3, 3), stride=(1, 1), padding=(2, 2), dilation=(2, 2), bias=False)
    (bn2): BatchNorm2d(256, eps=1e-05, momentum=0.1, affine=True, track_running_stats=True)
    (conv3): Conv2d(256, 1024, kernel_size=(1, 1), stride=(1, 1), bias=False)
    (bn3): BatchNorm2d(1024, eps=1e-05, momentum=0.1, affine=True, track_running_stats=True)
    (relu): ReLU(inplace=True)
  )
  (14): Bottleneck(
    (conv1): Conv2d(1024, 256, kernel_size=(1, 1), stride=(1, 1), bias=False)
    (bn1): BatchNorm2d(256, eps=1e-05, momentum=0.1, affine=True, track_running_stats=True)
    (conv2): Conv2d(256, 256, kernel_size=(3, 3), stride=(1, 1), padding=(2, 2), dilation=(2, 2), bias=False)
    (bn2): BatchNorm2d(256, eps=1e-05, momentum=0.1, affine=True, track_running_stats=True)
    (conv3): Conv2d(256, 1024, kernel_size=(1, 1), stride=(1, 1), bias=False)
    (bn3): BatchNorm2d(1024, eps=1e-05, momentum=0.1, affine=True, track_running_stats=True)
    (relu): ReLU(inplace=True)
  )
  (15): Bottleneck(
```

(conv1): Conv2d(1024, 256, kernel_size=(1, 1), stride=(1, 1), bias=False)

    (bn1): BatchNorm2d(256, eps=1e-05, momentum=0.1, affine=True, track_running_stats=True)

    (conv2): Conv2d(256, 256, kernel_size=(3, 3), stride=(1, 1), padding=(2, 2), dilation=(2, 2), bias=False)

    (bn2): BatchNorm2d(256, eps=1e-05, momentum=0.1, affine=True, track_running_stats=True)

    (conv3): Conv2d(256, 1024, kernel_size=(1, 1), stride=(1, 1), bias=False)

    (bn3): BatchNorm2d(1024, eps=1e-05, momentum=0.1, affine=True, track_running_stats=True)

    (relu): ReLU(inplace=True)

  )

  (16): Bottleneck(

    (conv1): Conv2d(1024, 256, kernel_size=(1, 1), stride=(1, 1), bias=False)

    (bn1): BatchNorm2d(256, eps=1e-05, momentum=0.1, affine=True, track_running_stats=True)

    (conv2): Conv2d(256, 256, kernel_size=(3, 3), stride=(1, 1), padding=(2, 2), dilation=(2, 2), bias=False)

    (bn2): BatchNorm2d(256, eps=1e-05, momentum=0.1, affine=True, track_running_stats=True)

    (conv3): Conv2d(256, 1024, kernel_size=(1, 1), stride=(1, 1), bias=False)

    (bn3): BatchNorm2d(1024, eps=1e-05, momentum=0.1, affine=True, track_running_stats=True)

    (relu): ReLU(inplace=True)

  )

  (17): Bottleneck(

    (conv1): Conv2d(1024, 256, kernel_size=(1, 1), stride=(1, 1), bias=False)

    (bn1): BatchNorm2d(256, eps=1e-05, momentum=0.1, affine=True, track_running_stats=True)

    (conv2): Conv2d(256, 256, kernel_size=(3, 3), stride=(1, 1), padding=(2, 2), dilation=(2, 2), bias=False)

(bn2): BatchNorm2d(256, eps=1e-05, momentum=0.1, affine=True, track_running_stats=True)

      (conv3): Conv2d(256, 1024, kernel_size=(1, 1), stride=(1, 1), bias=False)

      (bn3): BatchNorm2d(1024, eps=1e-05, momentum=0.1, affine=True, track_running_stats=True)

      (relu): ReLU(inplace=True)

    )

    (18): Bottleneck(

      (conv1): Conv2d(1024, 256, kernel_size=(1, 1), stride=(1, 1), bias=False)

      (bn1): BatchNorm2d(256, eps=1e-05, momentum=0.1, affine=True, track_running_stats=True)

      (conv2): Conv2d(256, 256, kernel_size=(3, 3), stride=(1, 1), padding=(2, 2), dilation=(2, 2), bias=False)

      (bn2): BatchNorm2d(256, eps=1e-05, momentum=0.1, affine=True, track_running_stats=True)

      (conv3): Conv2d(256, 1024, kernel_size=(1, 1), stride=(1, 1), bias=False)

      (bn3): BatchNorm2d(1024, eps=1e-05, momentum=0.1, affine=True, track_running_stats=True)

      (relu): ReLU(inplace=True)

    )

    (19): Bottleneck(

      (conv1): Conv2d(1024, 256, kernel_size=(1, 1), stride=(1, 1), bias=False)

      (bn1): BatchNorm2d(256, eps=1e-05, momentum=0.1, affine=True, track_running_stats=True)

      (conv2): Conv2d(256, 256, kernel_size=(3, 3), stride=(1, 1), padding=(2, 2), dilation=(2, 2), bias=False)

      (bn2): BatchNorm2d(256, eps=1e-05, momentum=0.1, affine=True, track_running_stats=True)

      (conv3): Conv2d(256, 1024, kernel_size=(1, 1), stride=(1, 1), bias=False)

      (bn3): BatchNorm2d(1024, eps=1e-05, momentum=0.1, affine=True, track_running_stats=True)

      (relu): ReLU(inplace=True)

```
    )
    (20): Bottleneck(
      (conv1): Conv2d(1024, 256, kernel_size=(1, 1), stride=(1, 1), bias=False)
      (bn1): BatchNorm2d(256, eps=1e-05, momentum=0.1, affine=True, track_running_stats=True)
      (conv2): Conv2d(256, 256, kernel_size=(3, 3), stride=(1, 1), padding=(2, 2), dilation=(2, 2), bias=False)
      (bn2): BatchNorm2d(256, eps=1e-05, momentum=0.1, affine=True, track_running_stats=True)
      (conv3): Conv2d(256, 1024, kernel_size=(1, 1), stride=(1, 1), bias=False)
      (bn3): BatchNorm2d(1024, eps=1e-05, momentum=0.1, affine=True, track_running_stats=True)
      (relu): ReLU(inplace=True)
    )
    (21): Bottleneck(
      (conv1): Conv2d(1024, 256, kernel_size=(1, 1), stride=(1, 1), bias=False)
      (bn1): BatchNorm2d(256, eps=1e-05, momentum=0.1, affine=True, track_running_stats=True)
      (conv2): Conv2d(256, 256, kernel_size=(3, 3), stride=(1, 1), padding=(2, 2), dilation=(2, 2), bias=False)
      (bn2): BatchNorm2d(256, eps=1e-05, momentum=0.1, affine=True, track_running_stats=True)
      (conv3): Conv2d(256, 1024, kernel_size=(1, 1), stride=(1, 1), bias=False)
      (bn3): BatchNorm2d(1024, eps=1e-05, momentum=0.1, affine=True, track_running_stats=True)
      (relu): ReLU(inplace=True)
    )
    (22): Bottleneck(
      (conv1): Conv2d(1024, 256, kernel_size=(1, 1), stride=(1, 1), bias=False)
      (bn1): BatchNorm2d(256, eps=1e-05, momentum=0.1, affine=True, track_running_stats=True)
```

(conv2): Conv2d(256, 256, kernel_size=(3, 3), stride=(1, 1), padding=(2, 2), dilation=(2, 2), bias=False)

            (bn2): BatchNorm2d(256, eps=1e-05, momentum=0.1, affine=True, track_running_stats=True)

            (conv3): Conv2d(256, 1024, kernel_size=(1, 1), stride=(1, 1), bias=False)

            (bn3): BatchNorm2d(1024, eps=1e-05, momentum=0.1, affine=True, track_running_stats=True)

            (relu): ReLU(inplace=True)

          )

        )

        (layer4): Sequential(

          (0): Bottleneck(

            (conv1): Conv2d(1024, 512, kernel_size=(1, 1), stride=(1, 1), bias=False)

            (bn1): BatchNorm2d(512, eps=1e-05, momentum=0.1, affine=True, track_running_stats=True)

            (conv2): Conv2d(512, 512, kernel_size=(3, 3), stride=(1, 1), padding=(2, 2), dilation=(2, 2), bias=False)

            (bn2): BatchNorm2d(512, eps=1e-05, momentum=0.1, affine=True, track_running_stats=True)

            (conv3): Conv2d(512, 2048, kernel_size=(1, 1), stride=(1, 1), bias=False)

            (bn3): BatchNorm2d(2048, eps=1e-05, momentum=0.1, affine=True, track_running_stats=True)

            (relu): ReLU(inplace=True)

            (downsample): Sequential(

              (0): Conv2d(1024, 2048, kernel_size=(1, 1), stride=(1, 1), bias=False)

              (1): BatchNorm2d(2048, eps=1e-05, momentum=0.1, affine=True, track_running_stats=True)

            )

          )

          (1): Bottleneck(

            (conv1): Conv2d(2048, 512, kernel_size=(1, 1), stride=(1, 1), bias=False)

(bn1): BatchNorm2d(512, eps=1e-05, momentum=0.1, affine=True, track_running_stats=True)

      (conv2): Conv2d(512, 512, kernel_size=(3, 3), stride=(1, 1), padding=(4, 4), dilation=(4, 4), bias=False)

      (bn2): BatchNorm2d(512, eps=1e-05, momentum=0.1, affine=True, track_running_stats=True)

      (conv3): Conv2d(512, 2048, kernel_size=(1, 1), stride=(1, 1), bias=False)

      (bn3): BatchNorm2d(2048, eps=1e-05, momentum=0.1, affine=True, track_running_stats=True)

      (relu): ReLU(inplace=True)
    )
    (2): Bottleneck(
      (conv1): Conv2d(2048, 512, kernel_size=(1, 1), stride=(1, 1), bias=False)

      (bn1): BatchNorm2d(512, eps=1e-05, momentum=0.1, affine=True, track_running_stats=True)

      (conv2): Conv2d(512, 512, kernel_size=(3, 3), stride=(1, 1), padding=(4, 4), dilation=(4, 4), bias=False)

      (bn2): BatchNorm2d(512, eps=1e-05, momentum=0.1, affine=True, track_running_stats=True)

      (conv3): Conv2d(512, 2048, kernel_size=(1, 1), stride=(1, 1), bias=False)

      (bn3): BatchNorm2d(2048, eps=1e-05, momentum=0.1, affine=True, track_running_stats=True)

      (relu): ReLU(inplace=True)
    )
   )
  )
  (classifier): FCNHead(
    (0): Conv2d(2048, 512, kernel_size=(3, 3), stride=(1, 1), padding=(1, 1), bias=False)
    (1): BatchNorm2d(512, eps=1e-05, momentum=0.1, affine=True, track_running_stats=True)
    (2): ReLU()
    (3): Dropout(p=0.1, inplace=False)

(4): Conv2d(512, 6, kernel_size=(1, 1), stride=(1, 1))
    )
  )
)